\documentclass[trackchanges]{aastex701}

\begin{document}

\newcommand{\pcm}{~cm$^{-2}$}	
\newcommand{\pcmm}{~cm$^{-3}$}	
\newcommand{\kms}{~km\,s$^{-1}$}

\title{A new  submillimeter Class I methanol maser at 345.903 GHz \\ near the massive protostar S255IR NIRS3}

\author[0009-0003-5504-9346]{Svetlana Salii}
\affiliation{Astronomical Obseratory,  Ural Federal University, 48 Kuibyshev str., Yekaterinburg 620026, Russia}
\email{svetlana.salii@urfu.ru}  

\author[orcid=0000-0003-2793-8229]{Igor Zinchenko} 
\affiliation{Federal Research Center A.V. Gaponov-Grekhov Institute of Applied Physics of the Russian Academy of Sciences, 46 Ul’yanov str., Nizhny Novgorod 603951, Russia}
\email{zin@ipfran.ru} 

\author[0000-0001-7575-5254]{Andrej Sobolev}
\affiliation{Xinjiang Astronomical Observatory, Chinese Academy of Sciences, Urumqi 830011, People’s Republic of China}
\email{andrej.sobolev@urfu.ru}   



\begin{abstract}
   We present results of further investigation of 
    radio emission from the remarkable molecular core S255IR-SMA1 harboring a $\sim20~{\mathrm{M}}_\odot$  young stellar object NIRS3 which exhibited a disk-mediated accretion burst in 2015. 
    Our study is based on observations of this object with ALMA in Band 7 with the longest baselines, which provide an angular resolution of $\sim$15~mas. 
    The obtained data reveal several spots in the bright and narrow methanol $16_1-15_2 \, A^{-}$ line emission at 345.903~GHz. These spots are located close to H$_2$O masers, knots in the jet and arc-like structures seen in submillimeter continuum. At the same time, they are far away from the peaks of the Class~II methanol maser spots. High brightness temperatures and small line widths of the 345.903~GHz line emission indicate that this emission has a maser nature. Association with the shock tracers tells that these masers belong to Class~I methanol maser category.
    Non-LTE analysis of the physical parameters which lead to population inversion of $16_1-15_2 \, A^{-}$ transition shows that some known rare Class I methanol maser transitions are also inverted under these conditions.
\end{abstract}


\section{Introduction} 

The maser effect observed in the emission of various molecules is an indicator of special conditions in the investigated areas. In our work, we consider methanol (CH$_3$OH) masers that are detected in star formation regions. Methanol masers are divided into two main classes: I and II. Class II methanol masers, observed in the immediate vicinity of young, massive stellar objects, are excited by radiation from these objects. The most famous maser transition is Class II, at the frequency of 6.7 GHz. Class I methanol masers are observed at a significant distance from the parent young stellar object \citep{2018IAUS..336..158V} in  areas where H$_2$O masers appear and are associated with shocks.

Our target is the actively studied area of massive star formation S255IR, at the distance of $\sim$1.8~kpc \citep{2016MNRAS.460..283B}. A massive protostellar object NIRS3 with a mass of $\sim20$~M$_\odot$ and an associated disk-outflow system have been discovered in the  S255IR-SMA1 core \citep{2015ApJ...810...10Z}. In the 0.9~mm continuum, a central bright source ($\sim850$~K) stands out  elongated along the jet, and two pairs of bright knots to the northeast (NE1 and~NE2) and southwest (SW1 and~SW2) from the central source  (see Fig. 1 in  \citet{2024A&A...692A.181Z}). In the vicinity of NIRS3, Class II methanol masers were detected, ranging from the brightest, at a frequency of 6.7 GHz, to  a weak  maser in the $18_{-3}-17_{-4} \, E$ transition at a frequency of 345.919 GHz first described by~\cite{2025MNRAS.543L...9Z} (the positions of the emission peaks in the corresponding line are marked with asterisks in Fig.~\ref{fig:continuum_and_345903_map}). In the wider surroundings of NIRS3, methanol masers of Class~I~\citep{2023ApJ...954L..25B} and H$_2$O masers at 22 GHz~\citep{Hirota2021,2023A&A...680A.110C} have been observed.
 
\section{Observations}

The observations used in this paper are described in \citet{2024A&A...692A.181Z}. Briefly, we carried out our observations with  ALMA on 2021 September 3 toward the S255IR SMA1 under project \#2019.1.00315.S. PI: S.-Y.~Liu. The phase center in the ICRS reference frame was Right Ascension (RA) = 06$^\mathrm{h}$12$^\mathrm{h}$54{\fs}013 and Declination (Dec.) = $+$17\degr59\arcmin23{\farcs}050. The synthesized beam of the continuum map was $\sim 19\times 13$~mas, which corresponds to $\sim 34\times 23$~AU at the distance of S255IR. The synthesized beams for the spectral line maps are almost the same. The spectral resolution was 1.13~MHz ($\sim$1\kms).

\section{Results}

According to the Cologne Database for Molecular Spectroscopy \citep[CDMS;][]{2001A&A...370L..49M}, there is a large number of methanol lines in our bands. The list of the observed methanol lines  in the disk can be seen in Table 1 in \citet{2025MNRAS.543L...9Z}. Apart from the Class II methanol maser lines considered in that paper, a  line at 345.903 GHz which corresponds to the transition $16_1-15_2\, A^-$ is presented in that table.  The line at 345.903 GHz is bright enough in the region S255IR SMA1, and in some regions considered in \citet{2025MNRAS.543L...9Z}, its emission is brighter than what can be obtained in the LTE model.  Therefore, we examine the emission of this line more carefully.

\begin{figure*}[ht!]
    \centering
    \includegraphics[width=0.9\linewidth]{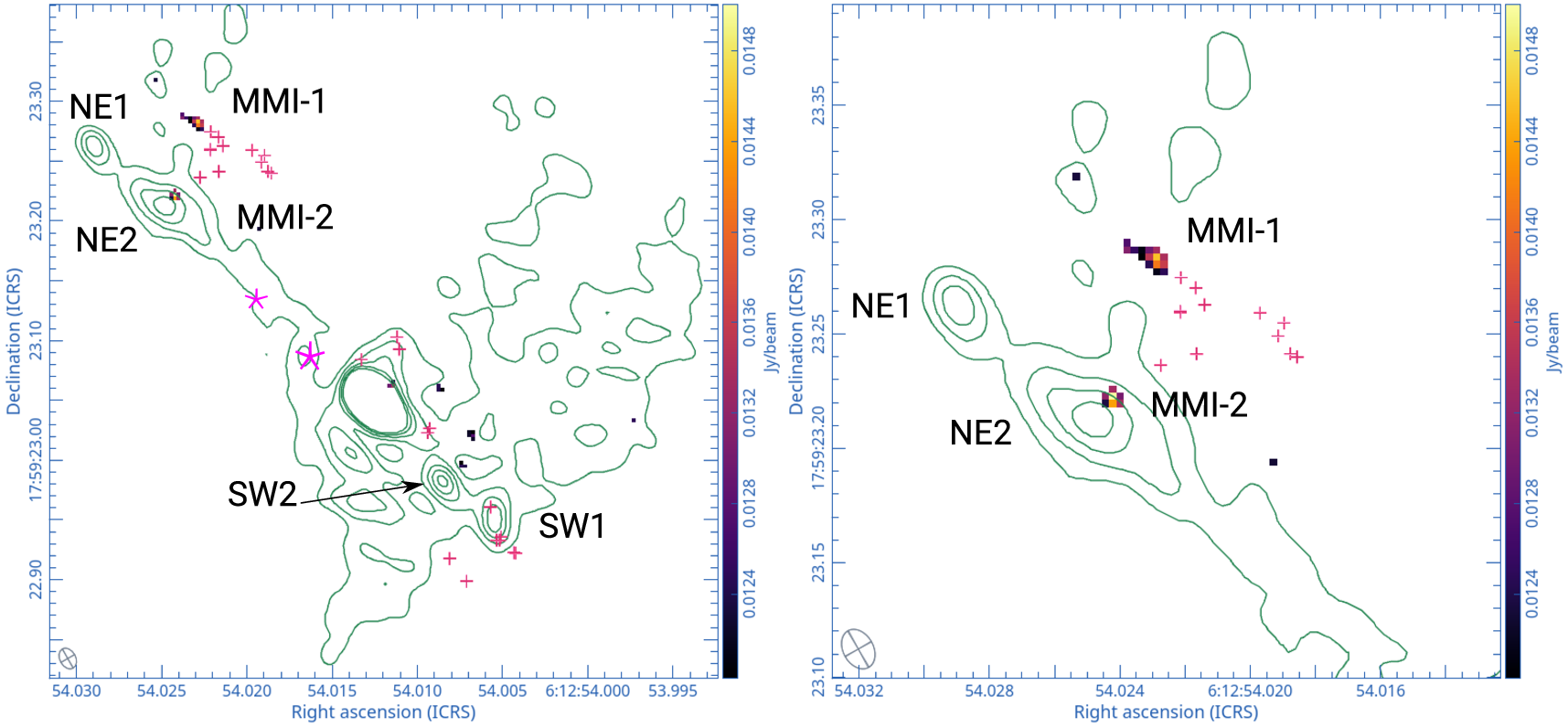}
    \caption{The map of the  peak  intensity in the CH$_3$OH $16_1-15_2\, A^-$ (345.903~GHz) line  with the lower intensity cutoff of 12~mJy\,beam$^{-1}$, which corresponds to approximately 510~K  (image), overlaid with contours of the 0.9~mm continuum emission \citep{2024A&A...692A.181Z}. The contour levels are 0.5 to 2.0 by 0.375~mJy\,beam$^{-1}$.  Positions of the brightest emission peaks of the $18_{-3}-17_{-4} E$ line (Class II methanol maser) from \citet{2025MNRAS.543L...9Z}  are marked as magenta asterisks.  H$_2$O masers positions from \citet{Hirota2021} are marked by red crosses. 
    The synthesized beam is plotted in the lower left corner.}
    \label{fig:continuum_and_345903_map}
\end{figure*}

The bright ($\ga510$~K) emission in the line at 345.903~GHz (Fig.~\ref{fig:continuum_and_345903_map}) is observed near the NE2, NE1 and SW2 knots in the red-shifted and blue-shifted lobes of the jet described in \citet{2024A&A...692A.181Z}. The brightest emission is observed near the NE2 knot (MMI-1 and MMI-2) along the arc-like structure seen in  the continuum at 0.9 mm.   The MMI-1 spot is slightly brighter and larger than MMI-2 and shows a bow-like structure. The other group of intensity peaks of the line is located near the SW2 knot and appears as a semi-ring structure. 

Remarkably, the intensity peaks of the line at 345.903~GHz are located far away from the peaks of the methanol Class II maser emissions (magenta asterisks in Fig. \ref{fig:continuum_and_345903_map}).  

\begin{figure}
    \centering
    \includegraphics[width=0.6\linewidth]
    {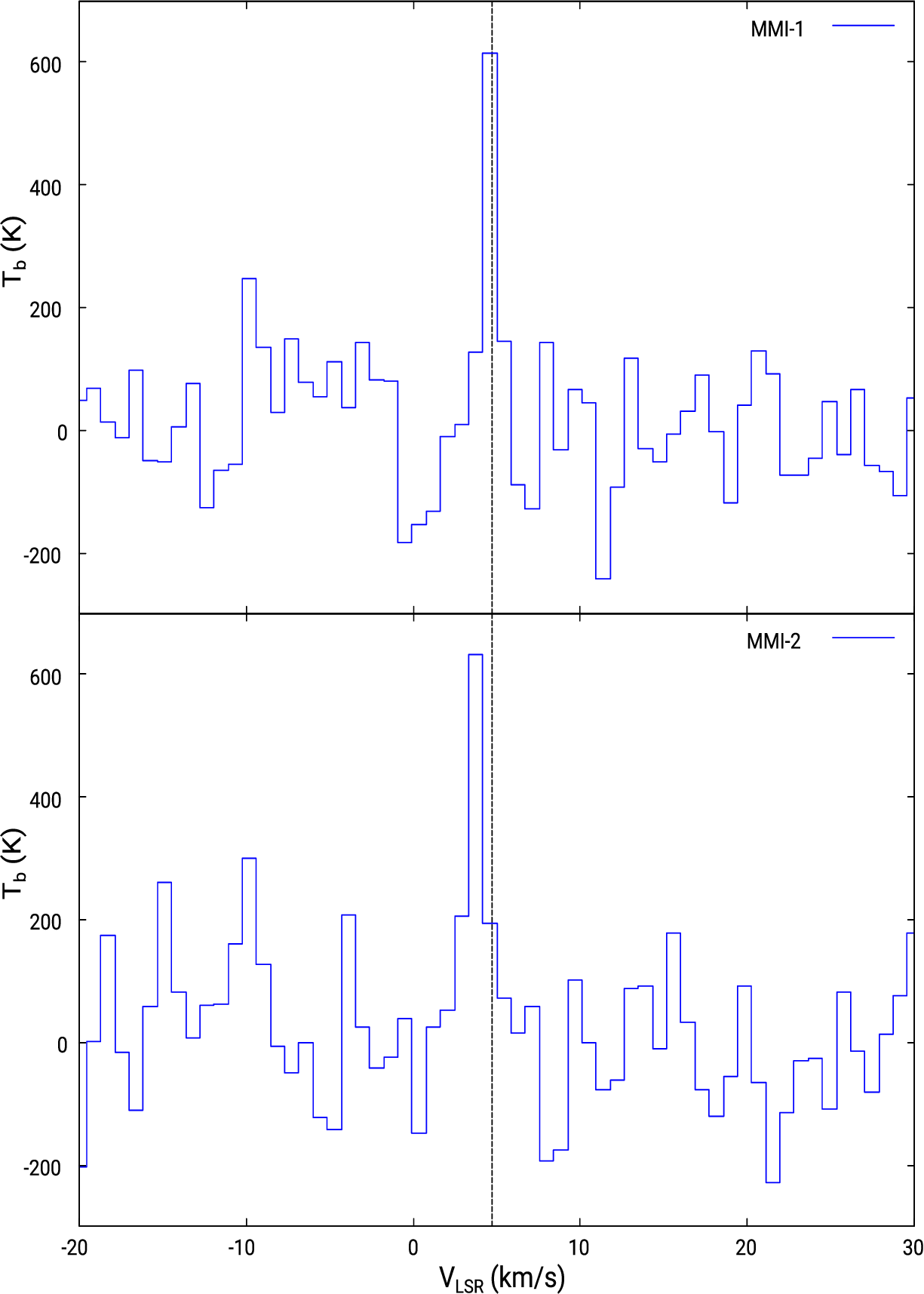}
    \caption{The spectra of the CH$_3$OH $16_1-15_2\, A^-$ (345.903~GHz) 
    toward the brightness peaks MMI-1 (R.A. = 06$^\mathrm{h}$12$^\mathrm{h}$54{\fs}0229, Dec. = $+$17\degr59\arcmin23{\farcs}28)(upper panel) and  MMI-2 (R.A. = 06$^\mathrm{h}$12$^\mathrm{h}$54{\fs}0242, Dec. = $+$17\degr59\arcmin23{\farcs}22 )(lower panel). Coordinates are in the ICRS frame. System velocity of source, 4.7 km/s is marked by dashed line. 
    }
    \label{fig:spectra_345903_345919_337135}
\end{figure}

In the regions MMI-1 and MMI-2,  the line is very narrow ($\la1.0$\kms) and its brightness reaches $\sim$600~K  (see Fig. \ref{fig:spectra_345903_345919_337135}).  

Despite the large number of methanol transitions in the observed bands, in the MMI-1 region  we confidently (up to $3\sigma$)  identified only  10 transitions. The list is presented in Table~\ref{tab:obs_methanol}. Among them, there are  Class II methanol maser transitions and torsionally excited transitions with excitation energies of up to 723 K. 

\begin{table}
\centering
\caption{Methanol transitions identified in our observations in the MMI-1 area with a high angular resolution. $E_{up}$ is the upper energy of a transition, $A_{ij}$ is Einstein A coefficient, according to CDMS \citep{2001A&A...370L..49M}.} 
\label{tab:obs_methanol}
\begin{tabular}{ccccc}
\hline
\hline
Frequency&  Transition     & $E_{u}$ &$\lg(A_E)$          & Comm. \\
 MHz    &            & K & $\lg( \rm{s}^{-1})$\\                           
\hline
336438.224 	&$14_{7}-15_{6}$ A$^{+},\, v_t=0$ & 488.2  &$-4.44$ &\\
336865.149	&$12_{1}-12_{0}$ A$^{-+},\, v_t=0$ & 197.1  &$-3.39$& MMII\\
337135.853  &$3_{3}- 4_{2}$ E $ v_t=0$         & 61.6   &$-4.80$&\\
337252.172  &$7_{3}- 6_{3}$ A$^{+,-},\, v_t=2$  & 722.8  &$-3.86$& \\ 
337297.484	&$7_{1}- 6_{1}$ A$^{+},\, v_t=1$  & 390.0  &$-3.78$&bl\\  
337463.703	&$7_{6}- 6_{6}$ E, $v_t=1$         & 533.0  &$-4.35$& \\
345903.916	&$16_{1}-15_{2}$ A$^{-},\, v_t=0$ & 332.6  &$-3.98$& MMI ?\\
346202.719	&$5_{4}- 6_{3}$ A$^{-},\, v_t=0$  & 115.2  &$-4.66$&bl\\  
346204.271	&$5_{4}- 6_{3}$ A$^{+},\, v_t=0$  & 115.2  &$-4.66$&bl\\  
349106.997	&$14_{1}-14_{0}$ A$^{-+},\, v_t=0$ & 260.2  &$-3.36$&MMII\\
\hline
\end{tabular}
\tablecomments{MMII --- transition is known as Class II methanol maser, MMI ? --- implied Class I methanol maser, \\ bl ---  blended lines.}
\end{table}

A difference in $V_{LSR}$ of about 0.8\kms\ in the 345.903 GHz line between the  MMI-1 ($4.6\pm0.1$\kms) and MMI-2 ($3.8\pm0.1$\kms)  positions is seen in Fig. \ref{fig:spectra_345903_345919_337135}. In the emission of the other lines, the difference is not evident. 

\section{Discussion}

The  transition under consideration, $16_1-15_2A^{-}$ (345.903 GHz) is a transition of the $J_1-(J-1)_2 A^{-}$ series. One transition of this series, $10_1-9_2 A^{-}$ at a frequency of 23.444 GHz, is known as a Class I methanol maser  \citep{2011MNRAS.413.2339V}. This Class I methanol maser was discovered during the H$_2$O southern Galactic Plane Survey.  The authors  classified this maser  as  a weak or rare Class I methanol maser and showed that such masers are excited in regions where  strong shocks are propagating and which have higher than average temperature and density. Therefore, such masers are too difficult to observe. Inasmuch as  our transition belongs to the same series  we assume that the observed emission at  345.903 GHz is also a Class I methanol maser. The opportunity for the maser effect in that transition was mentioned by \cite{2012IAUS..287..433V}.

Velocities, $V_{LRS}$, for regions MMI-1 and MMI-2  are close enough to the system velocity of 4.7\kms\ that was determined by CS emission. This is consistent with the velocity distribution of Class I methanol masers \citep[e.g.,][]{2006MNRAS.373..411V,2017MNRAS.471.3915J}.

It is known that Class I methanol masers  are associated with shocked material in the regions where jets or outflows are interacting with the ambient dense molecular gas \citep{1990ApJ...364..555P,2004ApJS..155..149K} and trace regular structures such as lines or arcs \citep{2017MNRAS.471.3915J}. The H$_2$O masers were found towards the same locations as Class I methanol maser sources by \cite{2017MNRAS.471.3915J,Kim_2019,Hirota2021}.   So we compared  the position of the peaks of emission in the line at 345.903 GHz with the positions of the H$_2$O maser features from \citet[][Table 2]{Hirota2021}. Superposition of H$_2$O maser features  with emission at 345.903~GHz is shown in Fig. \ref{fig:continuum_and_345903_map} (H$_2$O maser features are  red crosses). As can be clearly seen, the peaks at 345.903~GHz appear in the vicinity of the H$_2$O maser spots. Remarkably, the H$_2$O maser features with  the highest values of~proper motion are located near the region with the brightest emission at 345.903~GHz  MMI-1.

\begin{figure}
    \centering
    \includegraphics[width=0.98\linewidth]{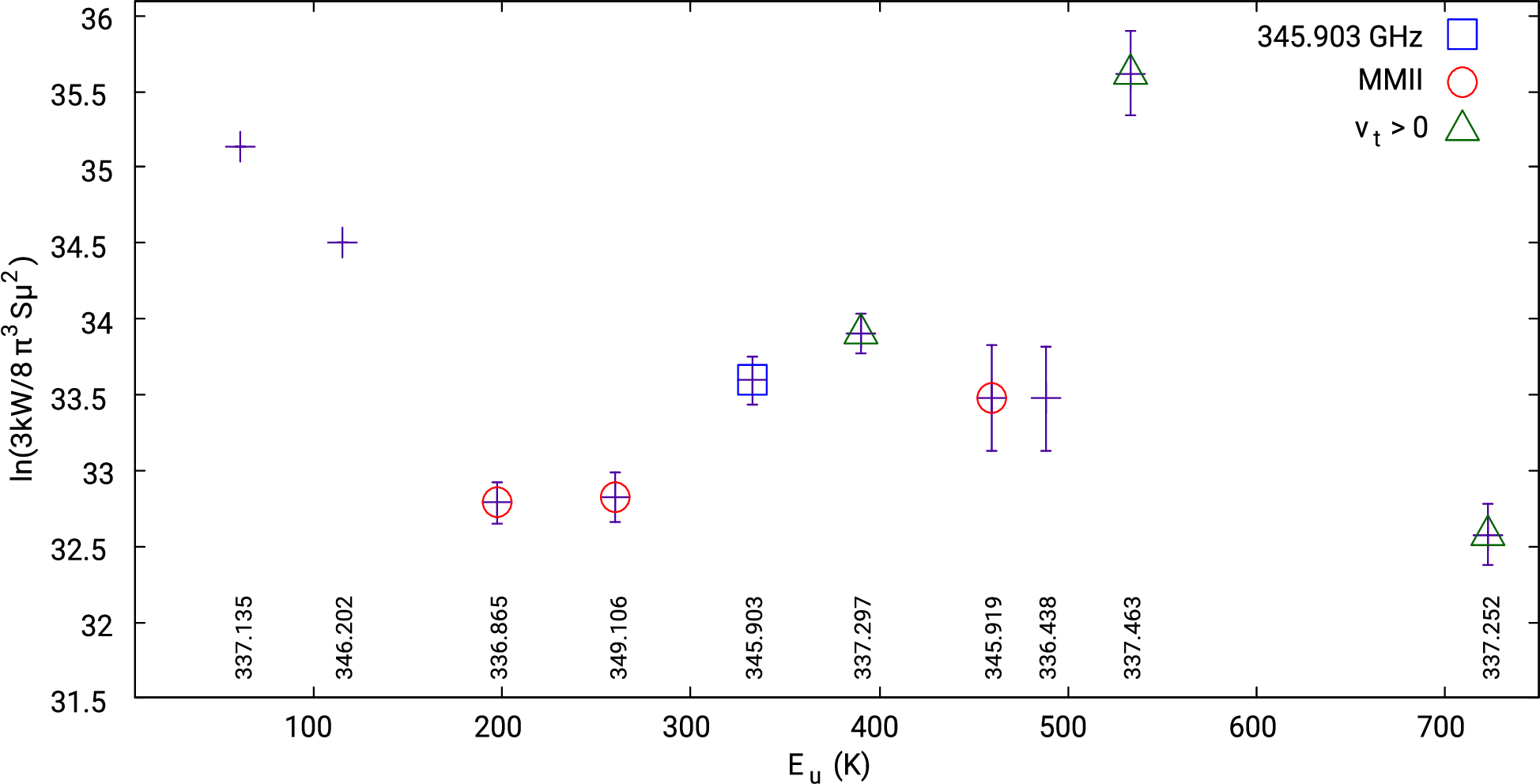}
    \caption{Rotation diagram for methanol lines in MMI-1.}
    \label{fig:rot_dia}
\end{figure}

For the region MMI-1 we made a rotation diagram (Fig. \ref{fig:rot_dia}). Here, one can see a very large deviation from the Boltzmann distribution that may be due to the very complex structure of the region and the fact that maser transitions,  transitions of the ground state, and torsionally excited states may arise in different parts of the source. Remarkably, the considered transition at 345.903 GHz  (blue square) is brighter than the potential Class II methanol maser transitions at 336.865 and 349.106 GHz (red circles) while its excitation energy is higher (332.6 K vs 197.1 and 260.2 K). This may be the result of the maser nature of the emission at 345.903 GHz.

Unfortunately, we could not find a good enough model to fit the observed transitions in a single set of parameters using the non-LTE approach described in \citet{2015ApJ...810...10Z}. We think  that this is due to  a small number (only three) quasi-thermal non-blended lines of the transitions in the ground state. Torsional excited and maser transitions are rather  excited under somewhat different physical conditions.

Apart from  the Class I methanol maser at 345.903 GHz,  \cite{2012IAUS..287..433V} predicted the possibility of the maser effect in the transition $11_0-10_1 A^+$ at 250.506 GHz, which is from the same series as the well known  Class I methanol maser at 44 GHz.  The  maser effect in this line in S255IR SMA1 was detected by  \cite{2023ApJ...954L..25B} at the southern filamentary structure  (out of the region considered here) in the velocity range of 4.5--5.7\kms\  (see  \citep[Fig. 4.]{2023ApJ...954L..25B}) that is coincident with the $V_{LSR}$ of the peak emission at 345.903 GHz ($4.7\pm0.2$\kms) in MMI-1.  Furthermore,  the  Class I methanol maser at 266.838 GHz ($5_2-4_1 E$). firstly detected by \cite{Chen_2019},  was detected  near the S255IR SMA1 NIRS3, but  out of the region considered here too  \citep{2023ApJ...954L..25B}.   

We estimate the  parameter intervals at which the transitions  at 250.506,  266.838 and 345.903 GHz are inverted (see Table \ref{tab:parameter_intervals}). As can be seen from the table, conditions for inversion of the transitions are similar, but not the same. We found out condition under which the lines at 250.506 and 266.838 GHz are quasi-thermal, according to \citep{2023ApJ...954L..25B},   whereas  the emission at 345.903 GHz may be maser. They correspond to kinetic temperature of the gas $T_k = 200-540$ K, gas number density $n_{H2} \sim 6 \cdot 10^{6}-2\cdot10^{8} \mathrm{cm}^{-3}$ and  methanol column density $N_{CH3OH} \sim 2\cdot10^{17}-10^{18}\,\, \mathrm{cm}^{-2}$. 

\begin{table*}     
\begin{center}
\caption{
        Modeled parameter intervals at which the transitions are inverted with a probability of 68 \%.  $T_k$~--- gas kinetic temperature,  $n_{H_2}$~--- gas number density,  $N_{CH_3OH}$~--- methanol column density. $E_{up}$~--- the upper energy of a transition, $A_{ij}$~--- Einstein A coefficient  
		}  
\label{tab:parameter_intervals}
\begin{tabular}{ccccccc}
\hline
\hline
Frequency	&  Transition                 &$E_{up}$&$A_{ij}$ & $T_k$     & $n_{H_2}$      & $N_{CH_3OH}$  \\
	GHz	    & 	                          & K      &s$^{-1}$ &  K        & cm$^{-3}$      & cm$^{-2}$ \\
\hline
   250.506  & $11_0-10_1 \, A^+ \, v_t=0$ &  153.1 & $8.9\cdot10^{-5}$ & $40-180$  & $(0.2-56)\cdot10^{6}$   &  $(1.0-200)\cdot10^{15}$  \\
   266.838  & $5_2-4_1 \, E \, v_t=0$     &   57.0 & $7.8\cdot10^{-5}$ & $80-200$  & $(1.0-36000)\cdot10^{3}$ & $(2.5-50)\cdot10^{15}$   \\
   345.903  & $16_1-15_2 \, A^- \, v_t=0$ &  332.6 & $1.0\cdot10^{-4}$ & $50-540$  & $(0.4-180)\cdot10^{6}$    & $(3.2-1000)\cdot10^{15}$  \\
\hline
\end{tabular}
\end{center}
\end{table*}  

This analysis lead us to the conclusion that the observed emission in the transition $16_{1}-15_{2}$ A$^{-},\, v_t=0$ at 345.903 GHz in the MMI-1 has a maser nature. It is the first detection of the maser effect in this transition.

\section{Conclusions}
 
The main results of this study can be summarized as follows:
\begin{itemize}
    \item[i] The bright and narrow emission of S255IR-SMA1 molecular core in the  $16_{1}-15_{2}$ A$^{-},\, v_t=0$ transition of methanol at 345.903 GHz is detected in the regions MMI-1, MMI-2, and some points near SW2.
    \item[ii] The emission at 345.903 GHz in the MMI-1 shows characteristics of maser emission: narrow line width ($< 1$~\kms) and high brightness temperature (up to 600 K). This emission is classified as Class I methanol maser one.
    \item[iii] Analysis under LVG approximation have shown that the observed in MMI-1 combination of maser emission in the 345.903 GHz transition and absence of inversion in 250.506 GHz and 266.838 GHz transitions is realized in the following range of physical parameters: $T_k = 200-540$ K, $n_{H2} \sim 6 \cdot 10^{6}-2\cdot10^{8}\,\, \mathrm{cm}^{-3}$ and $N_{CH3OH} \sim 2\cdot10^{17}-10^{18} \mathrm{cm}^{-2}$.
\end{itemize}
 
\begin{acknowledgments}
This work  was carried out with the financial support of the Ministry of Science and Higher Education of the Russian Federation (theme № FEUZ-2026-0012)  and the Russian Science Foundation grant No. 24-12-00153 (https://rscf.ru/en/project/24-12-00153/; data reduction and a part of the discussion by IIZ).
This paper makes use of the following ALMA data: ADS/JAO.ALMA \#2019.1.00315.S PI: S.-Y. Liu. ALMA is a partnership of ESO (representing its member states), NSF (USA), and NINS (Japan), together with NRC (Canada), MoST and ASIAA (Taiwan), and KASI (Republic of Korea), in cooperation with the Republic of Chile. The Joint ALMA Observatory is operated by ESO, AUI/NRAO, and NAOJ.
\end{acknowledgments}

\section*{Data Availability}
The {raw and pipeline processed} data are available in the ALMA archive {(ADS/JAO.ALMA \#2019.1.00315.S PI: S.-Y. Liu). The spectra and maps discussed in this paper are available by request from the corresponding author.}





%
\facilities{ALMA}

\software{CLASS software from the GILDAS packag (\url{http://www.iram.fr/IRAMFR/GILDAS})    }



\bibliography{s255ir-maser}{}
\bibliographystyle{aasjournalv7}



\end{document}